\documentclass[a4paper,11pt,centertags,psamsfonts]{amsart}
\usepackage{times,epsfig,amssymb}
\usepackage[mathcal]{euscript}

\DeclareSymbolFont{operators}   {OT1}{ptmcm}{m}{n}
\DeclareSymbolFont{letters}     {OML}{ptmcm}{m}{it}
\DeclareSymbolFont{symbols}     {OMS}{pzccm}{m}{n}
\DeclareSymbolFont{largesymbols}{OMX}{psycm}{m}{n}

\DeclareMathAlphabet{\mathsf}{OT1}{phv}{m}{n}
\DeclareMathAlphabet{\mathrm}{OT1}{ptm}{m}{n}

\DeclareSymbolFont{ER}{U}{eur}{m}{n}
\DeclareSymbolFont{SY}{U}{psy}{m}{n}

\DeclareMathSymbol{,}{\mathpunct}{SY}{'054}
\DeclareMathSymbol{.}{\mathpunct}{SY}{'056}
\DeclareMathSymbol{:}{\mathpunct}{SY}{'072}

\DeclareMathSymbol{(}{\mathopen}{SY}{'050}
\DeclareMathSymbol{)}{\mathclose}{SY}{'051}

\DeclareMathSymbol{+}{\mathbin}{SY}{'053}
\DeclareMathSymbol{-}{\mathbin}{SY}{'055}
\DeclareMathSymbol{=}{\mathbin}{SY}{'075}
\DeclareMathSymbol{<}{\mathbin}{SY}{'074}
\DeclareMathSymbol{>}{\mathbin}{SY}{'076}
\DeclareMathSymbol{\leq}{\mathbin}{SY}{'243}
\DeclareMathSymbol{\geq}{\mathbin}{SY}{'263}
\DeclareMathSymbol{\nneq}{\mathbin}{SY}{'271}
\DeclareMathSymbol{\in}{\mathbin}{SY}{'316}
\DeclareMathSymbol{\nnotin}{\mathbin}{SY}{'317}
\DeclareMathSymbol{\times}{\mathbin}{SY}{'264}
\DeclareMathSymbol{\pm}{\mathbin}{SY}{'261}
\DeclareMathSymbol{\subset}{\mathbin}{SY}{'314}
\DeclareMathSymbol{\supset}{\mathbin}{SY}{'311}
\DeclareMathSymbol{\subseteq}{\mathbin}{SY}{'315}
\DeclareMathSymbol{\supseteq}{\mathbin}{SY}{'312}

\DeclareMathSymbol{/}{\mathord}{SY}{'057}
\DeclareMathSymbol{\ast}{\mathord}{SY}{'052}
\DeclareMathSymbol{\perp}{\mathord}{SY}{'136}

\newcommand{\Z}{\mathbb{Z}}

\newcommand{\R}{\mathbb{R}}

\newcommand{\N}{\mathbb{N}}

\renewcommand{\P}{\mathbb{P}}
\newcommand{\E}{\mathbb{E}}

\newcommand{\EE}{\mathsf{E}}

\newcommand{\cJ}{\mathcal{J}}

\newcommand{\cQ}{\mathcal{Q}}

\newcommand{\supp}{{\ensuremath{\mathrm{supp}}}}

\newcommand{\tr}{\mathrm{tr}}

\newcommand{\meas}{\mathrm{meas}}

\newcommand{\sign}{\mathrm{sign}}

\newcommand{\fF}{\mathfrak{F}}

\DeclareMathOperator{\Rank}{\mathrm{Rank}}

\newcommand{\spec}{{\ensuremath{\rm spec}}}

\DeclareMathSymbol{\emptyset}{\mathord}{SY}{'306}
\DeclareMathSymbol{\oplus}{\mathord}{SY}{'305}

\newtheorem{theorem}{Theorem}{\bf}{\it}
\newtheorem{proposition}[theorem]{Proposition}{\bf}{\it}
{\bf}{\it}
{\bf}{\it}
{\it}{\rm}
\newtheorem{lemma}[theorem]{Lemma}{\bf}{\it}
{\it}{\rm}
{\bf}{\it}

\title[Regularity of the Density of Surface States]
{Regularity of the Density of Surface States}

\author[V. Kostrykin and R. Schrader]{Vadim Kostrykin \and Robert Schrader$^\ast$}

\address{Vadim Kostrykin\\
Fraunhofer-Institut f\"{u}r Lasertechnik\\ Steinbachstra{\ss}e 15, D-52074\\ Aachen,
Germany}
\email{kostrykin@t-online.de, kostrykin@ilt.fhg.de}

\address{Robert Schrader\\ Institut f\"{u}r
Theoretische Physik\\ Freie Universit\"{a}t Berlin, Arnimallee 14\\ D-14195 Berlin,
Germany}
\email{schrader@physik.fu-berlin.de}



\thanks{$^\ast$ R.S. supported in part by
DFG SFB 288 ``Differentialgeometrie und Quantenphysik''}

\date{November 10, 2000}

\begin{document}

\begin{abstract}
We prove that the integrated density of surface states of continuous or
discrete Anderson-type random Schr\"{o}dinger operators is a measurable locally
integrable function rather than a signed measure or a distribution. This
generalize our recent results on the existence of the integrated density of
surface states in the continuous case and those of A.\ Chahrour in the discrete
case. The proof uses the new $L^p$-bound on the spectral shift function
recently obtained by Combes, Hislop, and Nakamura. Also we provide a simple
proof of their result on the H\"{o}lder continuity of the integrated density of
bulk states.
\end{abstract}

\maketitle
\thispagestyle{empty}

\section{Introduction and Main Results}

Recently Combes, Hislop, and Nakamura \cite{Combes:Hislop:Nakamura} proved a
remarkable inequality for the $L^p$-norm of the spectral shift function. A
generalization of this inequality was then found by Hundertmark and Simon
\cite{Hundertmark:Simon}. As an application of this inequality Combes, Hislop,
and Nakamura prove H\"{o}lder continuity of the integrated density of states for a
wide class of random Schr\"{o}dinger operators. In this article, using the
$L^p$-bound, we will prove that the integrated density of surface states of
continuous or discrete Schr\"{o}dinger operators is a measurable, locally
integrable function thus extending results of
\cite{Englisch:Kirsch:Schroeder:Simon}, \cite{Chahrour:1999}, and
\cite{Kostrykin:Schrader:2000a}. Also we will provide a simple proof of the
H\"{o}lder continuity of the integrated density of (bulk) states for some Anderson
type models. It is based on the combination of the $L^p$-bound with the
Birman-Solomyak formula for the spectral shift function \cite{Birman:Solomyak}.
Although the proof of the H\"{o}lder continuity has already been given in
\cite{Combes:Hislop:Nakamura} and is based on the same ideas, our proof is
simpler for the model we consider.

So we first consider random Schr\"{o}dinger operators of the form $H_\omega = H_0 +
V_\omega$ on $L^2(\R^\nu)$, $\nu\geq 2$ with $H_0=-\Delta$ and $V_\omega$ being
the random potential of Anderson type centered near a hypersurface in $\R^\nu$.
More precisely we consider a decomposition $\Z^\nu = \Z^{\nu_1}\oplus
\Z^{\nu_2}$ with $\nu_1 +
\nu_2 =\nu$, $\nu_2\geq 1$ and introduce random potentials of the form
\begin{equation}\label{eq1}
V_\omega (x) = \sum_{\textbf{j}\in\Z^{\nu_1}} \alpha_{\textbf{j}}(\omega)
f(x-\textbf{j}),
\end{equation}
where $\alpha_j(\omega)$ is a sequence of random i.i.d.\ variables on a
probability space $(\Omega, \fF,\P)$ with common distribution $\kappa$, i.e.\
$\fF$ is a $\sigma$-algebra on $\Omega$, $\P$ a probability measure on
$(\Omega,\fF)$, and $\kappa(B)=\P\{\alpha_{\textbf{j}}\in B\}$ for any Borel
subset $B$ of $\R$. Let $\E$ denote the expectation with respect to $\P$. The
random variables $\{\alpha_{\mathbf{j}}(\omega)\}_{\mathbf{j}\in\Z^{\nu_1}}$
are supposed to form a stationary, metrically transitive random field, i.e.\
there are measure preserving ergodic transformations
$\{T_{\mathbf{j}}\}_{\mathbf{j}\in\Z^{\nu_1}}$ such that
$\alpha_{\mathbf{j}}(T_{\mathbf{k}}\omega)=\alpha_{\mathbf{j}-\mathbf{k}}(\omega)$
for all $\omega\in\Omega$. The single-site potential $f$ is supposed to be
supported in the unit cube $\Delta_0$ centered at the origin, $\supp f\subseteq
\Delta_0=[-1/2,1/2]^\nu$ and $f\in L^2(\R^\nu)$. Additionally if $\nu\geq 4$
the potential $f$ is supposed to belong to $L^r(\R^\nu)$ with some $r>\nu/2$.
Throughout this article the constant $r$ will be assumed to be fixed. Instead
of the integer lattice in (\ref{eq1}) we can alternatively consider an
arbitrary lattice as discussed in \cite{Kostrykin:Schrader:2000a}.

Finally we assume that $f$ is sign-definite, i.e.\ either $f>0$ or $f<0$ on
sets of positive Lebesgue measure. Below and without loss of generality further
we will consider the case $f\geq 0$ only since the case $f\leq 0$ is completely
similar. Also $\supp\ \kappa$ is supposed to be bounded, i.e.\ there is
$\alpha_->-\infty$ and $\alpha_+<\infty$ such that $\alpha_- \leq
\alpha_{\mathbf{j}}(\omega)\leq\alpha_+$ for all $\mathbf{j}\in\Z^{\nu_1}$
and all $\omega\in\Omega$. Under these conditions the operator $H_\omega=H_0 +
V_\omega$ defined in the form sense is self-adjoint on $\cQ(H_0)$ for all
$\omega\in\Omega$. The assumptions on $V_\omega$ can be relaxed by requiring
that the expectations of certain quantities are finite. The corresponding
modifications are obvious and we will not dwell on them.

Let $A$ and $C$ be bounded self-adjoint operators and let $C$ be trace class.
The spectral shift function $\xi(\cdot;A+C,A)\in L^1(\R)$ is defined by the
trace formula
\begin{equation}\label{trace}
\tr(\phi(A+C)-\phi(A))=\int_\R \phi'(\lambda)\xi(\lambda;A+C,A) d\lambda
\end{equation}
which is valid for a sufficiently wide class of continuous functions $\phi$ and
$\|\xi\|_{L^1(\R)}\leq \|C\|_{\cJ_1}$ where $\|\cdot\|_{\cJ_1}$ is the trace
norm. For relative trace class perturbations the spectral shift function can be
defined by means of the invariance principle (see e.g.\
\cite{Birman:Yafaev,Yafaev}). In particular if $A$ and $B$ are self-adjoint
possibly unbounded but bounded below with common domain of definition and if
$(B+a)^{-p}-(A+a)^{-p}$ is trace class for some $a>0$ and $p>1$ then
\begin{equation}\label{invariance}
\xi(\lambda;B,A)=-\xi((\lambda+a)^{-p}; (B+a)^{-p}, (A+a)^{-p}).
\end{equation}
It vanishes for all $\lambda<\inf\{\spec(B), \spec(A)\}$. A detailed account on
the theory of the spectral shift function can be found in the review
\cite{Birman:Yafaev} and in the book \cite{Yafaev}. For recent studies we refer
to \cite{Kostrykin:2000,Gesztesy:Makarov:99} and references therein. Recently
the spectral shift function found a number of applications in the theory of
random Schr\"{o}dinger operators \cite{Kostrykin:Schrader:1999},
\cite{Kostrykin:Schrader:2000a}, \cite{Kostrykin:Schrader:2000b},
\cite{Chahrour:1999}, \cite{Chahrour:2000}, \cite{Simon:95},
\cite{Nakamura:2000}, \cite{Combes:Hislop:Nakamura}.

Let $\Lambda\subset\R^{\nu_1}$ be a rectangular box $[a_1, b_1]\times \ldots
\times [a_{\nu_1}, b_{\nu_1}]$. We understand the limit
$\Lambda\rightarrow\infty$ in the sense that $a_i\rightarrow -\infty$ and
$b_i\rightarrow\infty$ for all $i=1,\ldots,\nu_1$. For an arbitrary box
$\Lambda$ we define
\begin{equation}\label{ran:pot:def1}
V_{\omega,\Lambda}(x) = \sum_{\substack{\mathbf{j}\in \Z^{\nu_1} \\
\mathbf{j}\in\Lambda}}
\alpha_{\mathbf{j}}(\omega) f(\cdot-\mathbf{j}).
\end{equation}
In \cite{Kostrykin:Schrader:2000a} we proved that for any $g\in C_0^1(\R)$ the
limit
\begin{equation*}
\lim_{\Lambda\rightarrow\infty} \frac{1}{\meas_{\nu_1}(\Lambda)} \int_\R
g(\lambda) \xi(\lambda; H_0 +V_{\omega,\Lambda}, H_0) d\lambda\ =:\ \mu(g)
\end{equation*}
exists almost surely and is non-random. The linear functional $\mu(g)$ is
related to the density of surface states $\mu_s(g)$ (see
\cite{Englisch:Kirsch:Schroeder:Simon}) such that $\mu_s(g)=\mu(g')$ (with $g'$
being the derivative of $g$), where
\begin{equation*}
\mu_s(g)= \lim_{\substack{\Lambda_1\rightarrow\infty\\\Lambda_2\rightarrow\infty}}
\frac{1}{\meas_{\nu_1}(\Lambda_1)}\tr\left[\chi_{\Lambda_1\times\Lambda_2}
(g(H_0+V_{\omega,\Lambda_1})-g(H_0))\right],\quad g\in C_0^2,
\end{equation*}
almost surely for arbitrary sequences of boxes $\Lambda_1\subset\R^{\nu_1}$,
$\Lambda_2\subset\R^{\nu_2}$ tending to infinity. Englisch, Kirsch, Schr\"{o}der,
and Simon \cite{Englisch:Kirsch:Schroeder:Simon} analyzed the surface states
occuring at the boundary between two Anderson-type crystals and proved that the
distribution induced by the functional $\mu_s(g)$ (i.e.\ the density of surface
states) is of order (at most) 3. This result applies almost verbatim also to
interactions of type \eqref{eq1}, so we have
\begin{equation*}
\mu_s(g)=\int_\R g(\lambda) N'_s(\lambda) d\lambda,\qquad g\in C_0^3(\R),
\end{equation*}
where the distribution $N_s(\lambda)$ of order at most 2 is called the
integrated density of surface states.

Using a slightly different approach from ours Chahrour \cite{Chahrour:1999}
constructed the functional $\mu(g)$ for the case of discrete Schr\"{o}dinger
operators and showed that the integrated density of surface states
$N_s(\lambda)$ is a distribution of order (at most) 1. In \cite{Chahrour:2000}
he proved that for discrete Schr\"{o}dinger operators with
\textit{nonrandom} periodic potentials $N_s(\lambda)$ is a measurable function.

Further in \cite{Kostrykin:Schrader:2000a} we proved that the functional
$\mu(g)$ induces a signed Borel measure $d\Xi(\lambda)$ such that for any $g\in
C_0(\R)$
\begin{equation*}
\mu(g)=\int_\R g(\lambda) d\Xi(\lambda).
\end{equation*}
This result implies that ``$N_s(\lambda) d\lambda$" is a $\sigma$-finite Borel
measure.

We will now extend this result and prove

\begin{theorem}\label{Thm1}
For continuous Schr\"{o}dinger operators $H_\omega = H_0 + V_\omega$ with
$V_\omega$ being defined by \eqref{eq1} the (signed) density of surface states
measure $d\Xi(\lambda)$ is Lebesgue absolutely continuous.
\end{theorem}

In the other words Theorem \ref{Thm1} states that the integrated density of
surface states $N_s(\lambda)$ is a measurable locally integrable function.

However, it remains unclear whether $N_s(\lambda)$ possesses further regularity
properties, e.g.\ whether it is a function of locally bounded variation such
that ``$d N_s(\lambda)$" defines a measure. In fact, it is difficult to control
the smoothness of $N_s(\lambda)$ since it may oscillate rapidly due to the
presence of alternating surface states and surface "holes".

The results of our article \cite{Kostrykin:Schrader:2000a} extend almost
verbatim (actually with several simplifications) to the case of discrete
Schr\"{o}dinger operators. More precisely we consider discrete Schr\"{o}dinger
operators with random potentials on a hypersurface,
\begin{equation}\label{Jacobi:1}
(h_\omega u)(\mathbf{n})=(h_0 u)(\mathbf{n}) +
\widetilde{V}_\omega(\mathbf{n}_1)
\delta(\mathbf{n}_2) u(\mathbf{n}),\quad
(h_0 u)(\mathbf{n})=\sum_{|\mathbf{j}|=1} u(\mathbf{n}-\mathbf{j}),\quad
\mathbf{n}=(\mathbf{n}_1, \mathbf{n}_2)\in \Z^{\nu_1}\oplus \Z^{\nu_2},
\end{equation}
where $\delta(\mathbf{n}_2)$ is the Kronecker symbol and
$\widetilde{V}_\omega(\mathbf{n}_1)$ a metrically transitive random field on
$\Z^{\nu_1}$.

We will prove the following analogue of Theorem \ref{Thm1}

\begin{theorem}\label{Thm2}
For discrete Schr\"{o}dinger operators of the form \eqref{Jacobi:1} the (signed)
density of surface states measure $d\Xi(\lambda)$ is Lebesgue absolutely
continuous.
\end{theorem}

The proofs of Theorems \ref{Thm1} and \ref{Thm2} will be given in Section
\ref{sec:2}. Section \ref{sec:3} plays a complementary role. Its aim is to give
a simple proof of the Combes-Hislop-Nakamura result on the H\"{o}lder continuity of
the integrated density of (bulk) states for some random Schr\"{o}dinger operators.
The proof is based on the combination of the Combes-Hislop-Nakamura $L^p$-bound
with the formula of Birman and Solomyak \cite{Birman:Solomyak}. This
combination is a generalization of Simon's spectral averaging method which was
used to prove Lipshitz continuity of the integrated density of bulk states
(Wegner's estimate) for some random Jacobi matrices \cite{Simon:95}.

\textbf{Acknowledgements.} We are indebted to J.M. Combes for useful discussions
and for sending us the preliminary version of the preprint preprint
\cite{Combes:Hislop:Nakamura}.

\section{Proofs of Theorems \ref{Thm1} and \ref{Thm2}}\label{sec:2}

The two main ingredients of our approach to prove Theorem \ref{Thm1} are the
Banach-Alauglu theorem (see e.g.\ \cite{Reed:Simon:1}) and the
Combes-Hislop-Nakamura $L^p$-bound for the spectral shift function
\cite{Combes:Hislop:Nakamura} (see also its generalization by Hundertmark and
Simon in \cite{Hundertmark:Simon}).

It is generally known that the discrete case is much easier to handle than the
continuous case. Indeed in the discrete case (Theorem \ref{Thm2}) we actually
do not need the $L^p$-bound for the spectral shift function and will use
instead a well-known bound for finite rank perturbations. We note also that in
the case of finite rank perturbations this bound is implied by the $L^p$-bound.

In the sequel we will use the following well-known lemma, which is a direct
consequence of the Banach-Alauglu theorem.

\begin{lemma}\label{Banach:Alauglu}
Let $1<q\leq \infty$. Let the sequence of real valued functions $f_n\in
L^q(a,b)$ satisfy
\begin{equation*}
\int\limits_a^b |f_n(\lambda)|^q d\lambda \leq C\quad (q<\infty)\qquad \textrm{or}\qquad
\sup_{\lambda\in(a,b)} |f_n(\lambda)|\leq C\quad (q=\infty)
\end{equation*}
uniformly in $n$ for some $C<\infty$. If the sequence $f_n(\lambda) d\lambda$
converges weakly to a signed measure $d\mu(\lambda)$ then this measure is
absolutely continuous.
\end{lemma}

For reader's convenience we recall the proof. By the Banach-Alauglu theorem we
can find a subsequence $f_{n(i)}$ of $f_n$ which converges in the weak-$\ast$
topology, i.e.\ there exists $f_\infty\in L^q(a,b)\subset L^1(a,b)$ such that
\begin{equation*}
\int f_{n(i)}(\lambda) g(\lambda) d\lambda \rightarrow \int f_\infty(\lambda) g(\lambda)
d\lambda,\qquad  g\in L^{q/(q-1)}(a,b).
\end{equation*}
Thus the measure $d\mu(\lambda)=f_\infty(\lambda) d\lambda$ is absolutely
continuous.

We start with the proof of Theorem \ref{Thm2} which is much easier than the one
for Theorem \ref{Thm1}. For an arbitrary rectangular box
$\Lambda\subset\R^{\nu_1}$ with integer-valued vertices we define
\begin{equation*}
V_{\omega,\Lambda}(\mathbf{n})=\begin{cases}\widetilde{V}_{\omega}(\mathbf{n}_1)
\delta(\mathbf{n}_2), & \mathbf{n}_1\in\Lambda,\\
0, & \textrm{otherwise.}
\end{cases}
\end{equation*}
Adopting the results of our article \cite{Kostrykin:Schrader:2000a} to the case
of discrete Schr\"{o}dinger operators we have

\begin{proposition}\label{Hilfslemma}
For any $g\in C_0^1(\R)$ the limit
\begin{equation*}
\lim_{\Lambda\rightarrow\infty} \frac{1}{\meas_{\nu_1}(\Lambda)} \int_\R g(\lambda)
\xi(\lambda; h_0+V_{\omega,\Lambda}, h_0) d\lambda
\end{equation*}
exists and defines a linear functional $\mu(g)$ on $C_0^1(\R)$. This functional
extends to all $g\in C_0(\R)$ with the representation
\begin{equation*}
\mu(g)=\int g(\lambda) d\Xi(\lambda)
\end{equation*}
and with $\Xi$ being a signed Borel measure.
\end{proposition}

The main idea behind the proof of Proposition \ref{Hilfslemma} is to consider
the random fields $V_{\omega}^+(\textbf{n})$ and $V_{\omega}^-(\textbf{n})$
such that $V_{\omega}^+(\textbf{n})=\max\{V_{\omega}(\textbf{n}),0\}$ and
$V_{\omega}^-(\textbf{n})=\min\{V_{\omega}(\textbf{n}),0\}$. It is
straightforward to see that $V_{\omega}^+(\textbf{n})$ and
$V_{\omega}^-(\textbf{n})$ are stationary, $\Z^{\nu_1}$-metrically transitive
random fields. By the chain rule for the spectral shift function we have
\begin{equation}\label{SSDens:8:eq}
\xi(\lambda;h_0+V_{\omega,\Lambda}, h_0) = \xi(\lambda; h_0+V_{\omega,\Lambda}^+
+V_{\omega,\Lambda}^-,
h_0+V_{\omega,\Lambda}^+)+\xi(\lambda;h_0+V_{\omega,\Lambda}^+,h_0).
\end{equation}
The first term on the r.h.s.\ of \eqref{SSDens:8:eq} is non-positive and the
second is non-negative. The next step is to prove the almost sure existence of
the limits
\begin{equation*}
\lim_{\Lambda\rightarrow\infty} \frac{1}{\meas_{\nu_1}(\Lambda)} \int_\R g(\lambda)
\xi(\lambda; h_0+V_{\omega,\Lambda}^+, h_0) d\lambda\ =:\ \mu^+(g)
\end{equation*}
and
\begin{equation*}
\lim_{\Lambda\rightarrow\infty} \frac{1}{\meas_{\nu_1}(\Lambda)} \int_\R g(\lambda)
\xi(\lambda;
h_0+V_{\omega,\Lambda}^- +V_{\omega,\Lambda}^+, h_0+V_{\omega,\Lambda}^+)
d\lambda\ =:\ \mu^-(g)
\end{equation*}
for all $g\in C_0^1(\R)$. But this follows from arguments used in
\cite{Kostrykin:Schrader:2000a} or \cite{Chahrour:1999}. The functionals
$\mu^\pm(g)$ are sign-definite. By the Riesz representation theorem they define
Borel measures $\Xi^\pm(\cdot)$. Moreover we have $\mu(g)=\mu^+(g)+\mu^-(g)$
and therefore $\Xi(\cdot)=\Xi^+(\cdot)+\Xi^-(\cdot)$, where $\Xi^+(\cdot)$ is a
positive and $\Xi^-(\cdot)$ a negative Borel measure.

Now we note that $V_{\omega,\Lambda}$ is a finite rank perturbation,
\begin{equation*}
\Rank V_{\omega,\Lambda} \leq \meas_{\nu_1}(\Lambda).
\end{equation*}
Therefore we have
\begin{equation*}
\frac{1}{\meas_{\nu_1}(\Lambda)}|\xi(\lambda; h_0+V_{\omega,\Lambda},
h_0)|\leq 1.
\end{equation*}
Applying Lemma \ref{Banach:Alauglu} with $q=\infty$ and using the fact that
$C_0^1$-functions are dense in $C_0$ from Proposition \ref{Hilfslemma} we
immediately obtain that the measure $d\Xi$ is absolutely continuous. This
completes the proof of Theorem \ref{Thm2}.

We turn to the proof of Theorem \ref{Thm1}. We have the following analogue of
Proposition \ref{Hilfslemma} (see \cite[Section
5.2]{Kostrykin:Schrader:2000a}):

\begin{proposition}\label{Hilfslemma:cont}
For any $g\in C_0^1(\R)$ the limit
\begin{equation*}
\lim_{\Lambda\rightarrow\infty} \frac{1}{\meas_{\nu_1}(\Lambda)} \int_\R g(\lambda)
\xi(\lambda; H_0+V_{\omega,\Lambda}, H_0) d\lambda
\end{equation*}
exists and defines a linear functional $\mu(g)$ on $C_0^1(\R)$. This functional
extends to all $g\in C_0(\R)$ and admits the representation
\begin{equation*}
\mu(g)=\int g(\lambda) d\Xi(\lambda)
\end{equation*}
with $\Xi$ being a signed Borel measure.
\end{proposition}

Let $s_j(T)$ denote the singular values of a compact operator $T$. For any $0<
p<\infty$ define the functional $T\mapsto |T|_p$ by
\begin{equation*}
|T|_{p}^p=\sum_j s_j(T)^p.
\end{equation*}
As well known, for $p\geq 1$ this functional defines a norm. The set of compact
operators $T$ with finite $|T|_p$ we denote by $\cJ_p$. In particular, $\cJ_1$
is the space of all trace class operators and $\cJ_2$ is the space of all
Hilbert-Schmidt operators. If $T_i\in
\cJ_{p_i}$, $0<p_i<\infty$, $i=1,2$ then $T_1 T_2\in \cJ_p$ with
$p^{-1}=p_1^{-1} + p_2^{-1}$ and
\begin{equation}\label{Seite4}
| T_1 T_2 |_p \leq |T_1|_{p_1} |T_2|_{p_2}.
\end{equation}
The proof of this inequality can be found in \cite[Corollary
11.11]{Birman:Solomyak:book} (actually there is a misprint there).

The proof of Theorem \ref{Thm1} heavily relies on the following lemma which is
due to Combes, Hislop and Nakamura \cite{Combes:Hislop:Nakamura}. A
generalization of this result can be found in \cite{Hundertmark:Simon}.

\begin{lemma}\label{lemma:lp:bound}
Let $A$ be a bounded self-adjoint operator on a separable Hilbert space. Let
the trace class operator $C$ be in $\cJ_{1/p}$ for some $1\leq p< \infty$. Then
\begin{equation*}
\| \xi(\cdot; A+C, A)\|_{L^p} \leq
|C|_{1/p}^{1/p}.
\end{equation*}
\end{lemma}

In the case $p=1$ this bound provides the well-known $L^1$-bound for the
spectral shift function, $\| \xi(\cdot; A+C, A)\|_{L^1} \leq |C|_1$. The case
$p=\infty$ is relevant in the case of finite rank perturbations, $\| \xi(\cdot;
A+C, A)\|_{L^\infty}\leq \Rank C$.

To proceed, we recall the definition of the Birman-Solomyak spaces $l^p(L^q)$,
$1\leq p,q\leq \infty$. They are the sets of all measurable functions
satisfying $\|\phi\|_{l^p(L^q)}<\infty$ with
\begin{equation*}
\|\phi\|_{l^p(L^q)} = \left(\sum_{\mathbf{j}\in\Z^\nu}\left[\int_{\Delta_0}
|\phi(x+\mathbf{j})|^q dx \right]^{p/q} \right)^{1/p}
\end{equation*}
and $\Delta_0$ being the unit cube centered at the origin.

We will write $R_{\omega,\Lambda}(-c)=(H_0+ V_{\omega,\Lambda}+c)^{-1}$ and
$R_0(-c)=(H_0+c)^{-1}$ for the resolvents of the operators
$H_0+V_{\omega,\Lambda}$ and $H_0$ respectively. With the above assumptions on
$f$ we prove

\begin{proposition}\label{lemma:1}
Let $k$ be an integer such that $k>(\nu-1)/2$ if $\nu\geq 4$ and $k=1$ if
$\nu\leq 3$. Let $c$ be a sufficiently large positive number. Then for any $p>
\nu/2(k+1)$ such that $p\leq 4$ if $\nu\leq 3$ and $p<2r$ if $\nu\geq 4$ the difference
$R_{\omega,\Lambda}(-c)^k
- R_0(-c)^k$ satisfies the inequality
\begin{equation}\label{lp:bound}
|R_{\omega,\Lambda}(-c)^k - R_0(-c)^k|_p
\leq C\ \meas_{\nu_1}(\Lambda)^{1/p}
\end{equation}
with $C$ being a constant independent of $\Lambda$ and $\omega$.
\end{proposition}

\noindent\textit{Remarks:} 1. The number $p$ can always be chosen to satisfy $p<1$.

2. A result of this type was already proved by Combes, Hislop, and Nakamura in
\cite{Combes:Hislop:Nakamura} (Proposition 5.1). The new ingredient in
Proposition \ref{lemma:1} is the volume dependence in the bound
\eqref{lp:bound}.

3. The fact that $R_{\omega,\Lambda}(-c)^k- R_0(-c)^k\in\cJ_1$ for
$k>(\nu-1)/2$, $\nu\geq 4$, and sufficiently large $c$ follows from Theorem
XI.12 of Reed-Simon \cite{Reed:Simon:3}.

For any measurable function $W$ we define $W^{1/2}$ by $W^{1/2}=\sign W\cdot
|W|^{1/2}$. For the proof of Proposition \ref{lemma:1} and again with the
assumptions on $f$ we need the following

\begin{lemma}\label{lem}
Let $k\geq 1/2$ and $p>\nu/2k$, $p\geq 1$. Moreover, let $p\leq 4$ if $\nu\leq
3$ and $p<2r$ if $\nu\geq 4$. Then there is a constant $C>0$ depending on
$\kappa$, $f$, $k$, and $\nu$ only such that
\begin{equation*}
| R_0(-c)^k |V_{\omega,\Lambda}|^{1/2}|_{p} \leq C\
\meas_{\nu_1}(\Lambda)^{1/p}\qquad \textrm{and}\qquad
| V_{\omega,\Lambda}^{1/2} R_0(-c)^k|_{p} \leq C\
\meas_{\nu_1}(\Lambda)^{1/p}.
\end{equation*}
\end{lemma}

\noindent\textit{Remark:} Note that both inequalities
\begin{eqnarray*}
\nu/2k < p < 2r & \textrm{if} & \nu\geq 4,\\
\nu/2k < p \leq 4 & \textrm{if} & \nu\leq 3.
\end{eqnarray*}
can always be satisfied. Indeed, for any $k\geq 1/2$ and $\nu\leq 3$ the
inequality $\nu/2k < 4$ holds. Since $2r >\nu$ for $\nu\geq 4$ and any $k\geq
1/2$ we have $\nu/2k < 2r$.

\begin{proof}
We consider the operator $R_0(-c)^k |V_{\omega,\Lambda}|^{1/2}$. The operator
$V_{\omega,\Lambda}^{1/2}R_0(-c)^k$ may be discussed similarly. Define the
function
\begin{equation}\label{g:def}
g(x)=(x^2+c)^{-k},\qquad x\in\R^\nu.
\end{equation}
It is easy to verify that $g\in L^q(\R^\nu)$ and $g\in l^q(L^2)$ for any
$q>\nu/2k$.

Suppose first that $\nu\leq 3$. From the assumption $f\in L^2(\R^\nu)$ and the
support property of $f$ it follows that $f\in L^1(\R^\nu)\cap L^2(\R^\nu)$ and
thus $f\in L^q(\R^\nu)$ for any $1\leq q\leq 2$. If $\nu\geq 4$ then from the
assumption that $f\in L^2(\R^\nu)\cap L^r(\R^\nu)$ for some $r>\nu/2$ and since
$f$ has compact support it follows that $f\in L^q(\R^\nu)$ for any $1\leq q<
r$.

For the case $1\leq p \leq 2$ we estimate as follows
\begin{eqnarray}\label{est1}
\| V_{\omega,\Lambda}^{1/2}\|_{l^p(L^2)}^p &=& \Big\|\sum_{\substack{\mathbf{j}\in \Z^{\nu_1}
\nonumber\\
\mathbf{j}\in\Lambda}}
|\alpha_{\mathbf{j}}(\omega)|^{1/2} f(\cdot-\mathbf{j})^{1/2}
\Big\|_{l^p(L^2)}^p\\ &=& \sum_{\substack{\mathbf{j}\in \Z^{\nu_1} \\
\mathbf{j}\in\Lambda}} \left[\int_{\Delta_0}
|\alpha_{\mathbf{j}}(\omega)| f(x) dx\right]^{p/2}\\ & = &
\sum_{\substack{\mathbf{j}\in \Z^{\nu_1} \\
\mathbf{j}\in\Lambda}} |\alpha_j(\omega)|^{p/2} \|f\|_{L^1}^{p/2}\ \leq\
C_1\ \meas_{\nu_1}(\Lambda)\nonumber
\end{eqnarray}
with $C_1$ being some constant depending on $\kappa$, $f$, $p$, and $\nu$ only.
Similarly for the case $2\leq p <\infty$ we have
\begin{eqnarray}\label{est2}
\| V_{\omega,\Lambda}^{1/2}\|_{L^p(\R^\nu)}^p &=&
\Big\|\sum_{\substack{\mathbf{j}\in \Z^{\nu_1} \\
\mathbf{j}\in\Lambda}}
|\alpha_{\mathbf{j}}(\omega)|^{1/2} f(\cdot-\mathbf{j})^{1/2}
\Big\|_{L^p(\R^\nu)}^p
\nonumber\\
&=& \sum_{\substack{\mathbf{j}\in \Z^{\nu_1} \\
\mathbf{j}\in\Lambda}} \int_{\Delta_0} |\alpha_j(\omega)|^{p/2} f(x)^{p/2} dx\\
&\leq & \sum_{\substack{\mathbf{j}\in \Z^{\nu_1} \\
\mathbf{j}\in\Lambda}} |\alpha_j(\omega)|^{p/2} \|f\|_{L^{p/2}(\R^\nu)}^{p/2}\ \leq\
C_2\ \meas_{\nu_1}(\Lambda)\nonumber
\end{eqnarray}
with $C_2$ depending again on $\kappa$, $f$, $p$, and $\nu$ only.

To estimate the norm of $R_0(-c)^k |V_{\omega,\Lambda}|^{1/2}$  we use the
Birman-Solomyak inequality \cite{Birman:Solomyak:69} (see also
\cite{Simon:book})
\begin{equation*}
| w\ g(-i\nabla)|_q \leq C_q\ \|w\|_{l^q(L^2)}\ \|g\|_{l^q(L^2)},\qquad 1\leq
q\leq 2
\end{equation*}
and the Seiler-Simon inequality (see \cite{Simon:book})
\begin{equation*}
| w\ g(-i\nabla)|_q \leq (2\pi)^{-\nu/q}\ \|w\|_{L^q}\ \|g\|_{L^q},\qquad 2\leq
q<\infty.
\end{equation*}
Setting $w=|V_{\omega,\Lambda}|^{1/2}$ and $g$ given by
\eqref{g:def} in these inequalities and then using the estimates \eqref{est1} and
\eqref{est2} proves the lemma.
\end{proof}

\begin{proof}[Proof of Proposition \ref{lemma:1}]
First we consider the case $\nu\leq 3$. By the resolvent equation
\begin{equation*}
R_{\omega,\Lambda}(-c)-R_0(-c) = - R_0(-c) |V_{\omega,\Lambda}|^{1/2} (I +
V_{\omega,\Lambda}^{1/2} R_0(-c) |V_{\omega,\Lambda}|^{1/2})^{-1}
V_{\omega,\Lambda}^{1/2} R_0(-c).
\end{equation*}
Since the operator norm of $(I + V_{\omega,\Lambda}^{1/2} R_0(-c)
|V_{\omega,\Lambda}|^{1/2})^{-1}$ is uniformly bounded, we obtain from
\eqref{Seite4} and Lemma \ref{lem}
\begin{equation*}
|R_{\omega,\Lambda}(-c)-R_0(-c)|_p \leq C\ |R_0(-c)
|V_{\omega,\Lambda}|^{1/2}|_{2p}\ |V_{\omega,\Lambda}^{1/2} R_0(-c)|_{2p} \leq
C\
\meas_{\nu_1}(\Lambda)^{1/p}.
\end{equation*}

We turn to the case $\nu\geq 4$. For any $k\in\N$ we have
\begin{eqnarray*}
R_{\omega,\Lambda}(-c)^k - R_0(-c)^k &=&
\frac{(-1)^{k-1}}{(k-1)!}\frac{d^{k-1}}{d c^{k-1}}\left[ R_{\omega,\Lambda}(-c)
 - R_0(-c)\right]\\
 &=& \sum_{\substack{l+m+n=k-1 \\l,m,n\in\N\cup\{0\} }} c_{lnm}\ R_0(-c)^{l+1}
 |V_{\omega,\Lambda}|^{1/2}\ K_m\  V_{\omega,\Lambda}^{1/2} R_0(-c)^{n+1},
\end{eqnarray*}
with some coefficients $c_{lnm}$. The operators $K_m$ are given by
\begin{equation}\label{Km:def}
K_m = \frac{d^m}{d c^m} (I+V_{\omega,\Lambda}^{1/2}
 R_0(-c) |V_{\omega,\Lambda}|^{1/2})^{-1},\quad m\in\N\cup\{0\}.
\end{equation}
Applying Lemma \ref{lem} we obtain
\begin{eqnarray*}
R_0(-c)^{l+1} |V_{\omega,\Lambda}|^{1/2} \in\cJ_{p_1}\quad & \mathrm{for}
&\quad p_1>\frac{\nu}{2(l+1)},\\ V_{\omega,\Lambda}^{1/2} R_0(-c)^{n+1}
\in\cJ_{p_2}\quad & \mathrm{for} & \quad p_2>\frac{\nu}{2(n+1)}
\end{eqnarray*}
with
\begin{eqnarray*}
|R_0(-c)^{l+1} |V_{\omega,\Lambda}|^{1/2}|_{p_1}& \leq & C\
\meas_{\nu_1}(\Lambda)^{1/p_1},\\
\\ |V_{\omega,\Lambda}^{1/2} R_0(-c)^{n+1}|_{p_2} & \leq &
C\ \meas_{\nu_1}(\Lambda)^{1/p_2}.
\end{eqnarray*}

We turn to the discussion of the operators $K_m$. Obviously we have
\begin{equation}\label{Ableit}
\begin{split}
&\frac{d}{dc} (I+V_{\omega,\Lambda}^{1/2}
 R_0(-c) |V_{\omega,\Lambda}|^{1/2})^{-1}\\ & = (I+V_{\omega,\Lambda}^{1/2}
 R_0(-c) |V_{\omega,\Lambda}|^{1/2})^{-1}\ V_{\omega,\Lambda}^{1/2}
 R_0(-c)^2 |V_{\omega,\Lambda}|^{1/2}\ (I+V_{\omega,\Lambda}^{1/2}
 R_0(-c) |V_{\omega,\Lambda}|^{1/2})^{-1}.
 \end{split}
\end{equation}
Let $m,i\in\N$, $i\leq m$ be given. By $M_{m,i}$ we denote the set of all
multiindices $\underline{m}=(m_1,\ldots,m_i)$ with $m_1,\ldots,m_i\in\N$
satisfying the following conditions
\begin{eqnarray*}
&& m_1,\ \ldots,\ m_i\ \geq 2,\\  && m_1\ +
\ \ldots\ +\ m_i = m+i.
\end{eqnarray*}
Applying the formula \eqref{Ableit} to \eqref{Km:def} recursively we obtain
that for any $m\geq 1$ the operator $K_m$ can be represented in the form
\begin{equation}\label{Km:repr}
\begin{split}
K_m &= \sum_{i=1}^m \sum_{\underline{m}\in M_{mi}} c_{mi\underline{m}}\
(I+V_{\omega,\Lambda}^{1/2}
 R_0(-c) |V_{\omega,\Lambda}|^{1/2})^{-1} \\
 & \qquad\cdot \prod_{j=1}^i V_{\omega,\Lambda}^{1/2}\ R_0(-c)^{m_j}\
 |V_{\omega,\Lambda}|^{1/2}\
(I+V_{\omega,\Lambda}^{1/2}
 R_0(-c) |V_{\omega,\Lambda}|^{1/2})^{-1}
\end{split}
\end{equation}
with $c_{mi\underline{m}}$ being some real numbers.

We prove now that for any $m\geq 1$
\begin{equation}\label{Ziel1}
K_m\in \cJ_{p_3},\quad |K_m|_{p_3}\leq C\ \meas_{\nu_1}(\Lambda)^{1/p_3}\quad
\textrm{with}\quad p_3>\frac{\nu}{2(m+1)}.
\end{equation}
From Lemma \ref{lem} and using the inequality \eqref{Seite4} we obtain
\begin{equation*}
V_{\omega,\Lambda}^{1/2} R_0(-c)^{m_j} |V_{\omega,\Lambda}|^{1/2} \in
\cJ_{q_j}\quad
\textrm{with}\quad q_j>\frac{\nu}{2m_j}
\end{equation*}
and $q_j<4$ for $\nu\leq 3$ and $q_j < 2r$ for $\nu\geq 4$. Moreover the
inequality
\begin{equation*}
\left| V_{\omega,\Lambda}^{1/2}\ R_0(-c)^{m_j}\ |V_{\omega,\Lambda}|^{1/2} \right|_{q_j}
\leq C\ \meas_{\nu_1}(\Lambda)^{1/m_j}
\end{equation*}
holds. Thus
\begin{equation*}
\prod_{j=1}^i V_{\omega,\Lambda}^{1/2}\ R_0(-c)^{m_j}\ |V_{\omega,\Lambda}|^{1/2}
(I + V_{\omega,\Lambda}^{1/2}
 R_0(-c) |V_{\omega,\Lambda}|^{1/2})^{-1} \in \cJ_{\widetilde{q}_i}
\end{equation*}
with
\begin{equation*}
\widetilde{q}_i > \frac{\nu}{2\sum_{j=1}^i m_j} = \frac{\nu}{2(m+i)}
\end{equation*}
and such that $\widetilde{q}_i < 4$ for $\nu\leq 3$ and $\widetilde{q}_i < 2r$
for $\nu\geq 4$. Moreover the estimate
\begin{equation*}
\Big| \prod_{j=1}^i V_{\omega,\Lambda}^{1/2}\ R_0(-c)^{m_j}\ |V_{\omega,\Lambda}|^{1/2}\
(I + V_{\omega,\Lambda}^{1/2}
 R_0(-c) |V_{\omega,\Lambda}|^{1/2})^{-1}
 \Big|_{\widetilde{q}_i} \leq C\ \meas_{\nu_1}(\Lambda)^{1/\widetilde{q}_i}
\end{equation*}
holds. In the equation \eqref{Km:repr} the worst case occurs for $i=1$. Thus
the estimate \eqref{Ziel1} is proved.

From Lemma \ref{Seite4} it now follows that
\begin{equation*}
R_{\omega,\Lambda}(-c)^k - R_0(-c)^k\in\cJ_p
\end{equation*}
with $p$ satisfying
\begin{equation*}
\frac{1}{p}=\frac{1}{p_1}+\frac{1}{p_2}+\frac{1}{p_3}<\frac{2(l+1)+2(n+1)+2m+2}{\nu}
=\frac{2(k+2)}{\nu}
\end{equation*}
and the estimate \eqref{lp:bound} holds. Since the inequality $p>\nu/(2(k+2))$
is satisfied with any $p>\nu/(2(k+1))$ this completes the proof.
\end{proof}

Now we are in the position to complete the proof of Theorem \ref{Thm1}. We
choose some $k>(\nu-1)/2$ if $\nu\geq 4$ and set $k=1$ if $\nu\leq 3$. Fix some
$p$ satisfying $1>p>\nu/2(k+1)$. Consider an arbitrary interval $(a,b)$ of the
real line. Using the invariance principle for the spectral shift function
\eqref{invariance} we estimate
\begin{eqnarray*}
\lefteqn{\int_a^b \left|\frac{\xi(\lambda; H_0+ V_{\omega,\Lambda}, H_0)}
{\meas_{\nu_1}(\Lambda)}\right|^{1/p} d\lambda}\\ &\leq &
\left(\meas_{\nu_1}(\Lambda)\right)^{-1/p}
\int_a^b \left|\xi((\lambda+c)^{-k}; R_{\omega,\Lambda}(-c)^k, R_0(-c)^k)\right|^{1/p}
d\lambda\\ &=& \left(\meas_{\nu_1}(\Lambda)\right)^{-1/p} k^{-1}
\int_{(b+c)^{-k}}^{(a+c)^{-k}}
\left|\xi(t; R_{\omega,\Lambda}(-c)^k, R_0(-c)^k)\right|^{1/p} t^{-1/k-1} dt\\
&\leq & \left(\meas_{\nu_1}(\Lambda)\right)^{-1/p} k^{-1} (b+c)^{k+1}
\int_\R
\left|\xi(t; R_{\omega,\Lambda}(-c)^k, R_0(-c)^k)\right|^{1/p} dt.
\end{eqnarray*}
Now applying Lemma \ref{lemma:lp:bound} we get
\begin{equation*}
\int_a^b \left(\frac{\xi(\lambda; H_0+ V_{\omega,\Lambda}, H_0)}
{\meas_{\nu_1}(\Lambda)}\right)^{1/p} d\lambda \leq C\
\left(\meas_{\nu_1}(\Lambda)\right)^{-1/p} |R_{\omega,\Lambda}(-c)^k - R_0(-c)^k|_p.
\end{equation*}
By Proposition \ref{lemma:1} the r.h.s.\ of this inequality is bounded
uniformly in $\Lambda$ and $\omega\in\Omega$. Thus Lemma \ref{Banach:Alauglu}
with $q=p$ proves the absolute continuity of $d\Xi$.

\section{H\"{o}lder Continuity of the Integrated Density of Bulk States}\label{sec:3}

Here we give a simple proof of the H\"{o}lder continuity of the integrated density
of bulk states for some random Schr\"{o}dinger operators based on the new
$L^p$-bound of Combes, Hislop, and Nakamura and on the formula of Birman and
Solomyak \cite{Birman:Solomyak}.

To be concrete, we consider the Holden-Martinelli model, where the single-site
potential is the characteristic function of the unit cube,
\begin{equation*}
H_\omega = -\Delta + \sum_{\mathbf{j}\in\Z^\nu} \alpha_{\mathbf{j}}(\omega)\
\chi(\cdot-j)\qquad \textrm{on}\quad L^2(\R^\nu).
\end{equation*}
The distribution $\kappa$ will be supposed to be absolutely continuous,
$d\kappa = p(\alpha)d\alpha$, and compactly supported, i.e.\ $\supp p\subseteq
[\alpha_-,\alpha_+]$. The integrated density of states has the following
representation (see \cite{Pastur:Figotin})
\begin{equation*}
N(\lambda) = \E\left\{\tr \left(\chi\
\EE_{H_\omega}((-\infty,\lambda))\ \chi\right)
\right\},
\end{equation*}
where $\EE_{H_\omega}$ denotes the spectral projection corresponding to
$H_\omega$. Lipshitz continuity of $N(\lambda)$ was proved in
\cite{Combes:Hislop:1994}.

With $I=(\lambda_1, \lambda_2)$ and $A_\omega=H_\omega|_{\alpha_0(\omega)=0}$
we consider
\begin{eqnarray}\label{zwischen}
N(\lambda_2)-N(\lambda_1)&=& \E\left\{\tr\left(\chi\
\EE_{H_\omega}(I)\ \chi\right)\right\}\nonumber\\ &=&
\E\left\{\int_{\alpha_-}^{\alpha_+} d\alpha\ p(\alpha)\ \tr(\chi\
\EE_{A_\omega+\alpha\chi}(I)\ \chi)\right\}\nonumber\\
&\leq & \|p\|_\infty \E\left\{\int_{\alpha_-}^{\alpha_+} d\alpha\
\tr\left(
\chi\ \EE_{A_\omega+\alpha\chi}(I)\ \chi\right)\right\}.
\end{eqnarray}
Now we will use the Birman-Solomyak formula \cite{Birman:Solomyak} (see also
\cite{Simon:98}, \cite{Gesztesy:Makarov:Motovilov}). The present formulation is
from \cite{Simon:98}. Let
\begin{equation*}
L^p_{\mathrm{unif,loc}}(\R^\nu):=\left\{\phi\Big|\ \sup_x
\int_{|x-y|\leq 1} |\phi(y)|^p dy < \infty\right\},\qquad p>\nu/2.
\end{equation*}

\begin{theorem}
Let $A=-\Delta+W$ with $W\in L^p_{\mathrm{unif,loc}}(\R^\nu)$ and $V\in
l^1(L^2)$, $V\geq 0$. For any compact interval $I\subset\R$ the following
relation is valid
\begin{equation*}
\int_I \xi(\lambda; A+\alpha_- V, A+\alpha_+ V) d\lambda = \int_{\alpha_-}^{\alpha_+}
\tr\left(V^{1/2} \EE_{A+sV}(I) V^{1/2} \right) ds.
\end{equation*}
\end{theorem}

Applying this theorem to \eqref{zwischen} we obtain
\begin{equation*}
N(\lambda_2)-N(\lambda_1)\leq \|p\|_\infty
\E\left\{\int_{\lambda_1}^{\lambda_2} \xi(\lambda; A_\omega + \alpha_+ \chi,
A_\omega + \alpha_-
\chi) d\lambda\right\}.
\end{equation*}
From the H\"{o}lder inequality it follows that for any $p>1$ and any
$\omega\in\Omega$
\begin{eqnarray*}
\int_{\lambda_1}^{\lambda_2} |\xi(\lambda; A_\omega + \alpha_+ \chi, A_\omega+ \alpha_-
\chi)| d\lambda \leq \left(\int_{\lambda_1}^{\lambda_2} \xi(\lambda; A_\omega +
\alpha_+ \chi, A_\omega + \alpha_- \chi)^p d\lambda\right)^{1/p}
|\lambda_2-\lambda_1|^{\frac{p-1}{p}}
\end{eqnarray*}
Choose some $k>(\nu-1)/2$ if $\nu\geq 4$ and $k\geq 1$ if $\nu\leq 3$ so large
that $k>\nu/2p - 1$. By the invariance principle for the spectral shift
function
\eqref{invariance} for any $\omega\in\Omega$ we have
\begin{equation}\label{X}
\begin{split}
&\int_a^b \xi(\lambda; A_\omega +
\alpha_+ \chi, A_\omega + \alpha_- \chi)^p d\lambda\\
&\quad = \int_a^b \left|\xi((\lambda+c)^{-k}; (A_\omega + \alpha_+ \chi +
c)^{-k}, (A_\omega + \alpha_- \chi + c)^{-k})
\right|^p d\lambda\\
&\quad = k^{-1} \int_{(b+c)^{-k}}^{(a+c)^{-k}} \left|\xi(t; (A_\omega +
\alpha_+
\chi + c)^{-k}, (A_\omega + \alpha_- \chi + c)^{-k})\right|^p t^{-1/k-1} dt\\
&\quad\leq k^{-1}(b+c)^{k+1}\int_\R \left|\xi(t; (A_\omega + \alpha_+
\chi + c)^{-k}, (A_\omega + \alpha_- \chi + c)^{-k})\right|^p dt.
\end{split}
\end{equation}
By Proposition 5.1 of \cite{Combes:Hislop:Nakamura} we have that
\begin{equation}\label{incl}
(A_\omega +
\alpha_+\chi +c)^{-k}- (A_\omega + \alpha_-\chi +c)^{-k}\in \cJ_{1/p}.
\end{equation}
Alternatively we can use our Proposition \ref{lemma:1}. For instance, for $k=1$
we have
\begin{equation*}
\begin{split}
&(A_\omega +
\alpha_+\chi +c)^{-1}- (A_\omega + \alpha_-\chi +c)^{-1} = \left[
I+(H_0+\alpha_- \chi + c)^{-1} V_\omega\big|_{\alpha_0(\omega)=0}\right]^{-1}\\
&\cdot \left\{(H_0+\alpha_+ \chi+c)^{-1}-(H_0+\alpha_- \chi+c)^{-1}\right\}
\left[
I-V_\omega\big|_{\alpha_0(\omega)=0}\left(A_\omega + \alpha_+ \chi +
c\right)^{-1}
\right].
\end{split}
\end{equation*}
Since for sufficiently large $c$ the first and the last factors on the r.h.s.\
of this equality are bounded uniformly in $\omega\in\Omega$, the relation
\eqref{incl} follows from Proposition \ref{lemma:1}. Since $\chi\in L^\infty$
the additional restrictions $p\leq 4$ if $\nu\leq 3$ and $p<2r$ if $\nu\geq 4$
can be omitted.

Thus from Lemma \ref{lemma:lp:bound} it follows that the l.h.s.\ of \eqref{X}
is bounded by a constant $C>0$ uniformly in $\omega\in\Omega$. Finally this
leads to the estimate
\begin{equation*}
N(\lambda_2) - N(\lambda_1) \leq C_p |\lambda_2 -\lambda_1|^{\frac{p-1}{p}}
\end{equation*}
for any $p>1$, which proves the H\"{o}lder continuity of the integrated density of
states. We can apply similar arguments to models more general than the
Holden-Martinelli model. We will not dwell on this here.


\end{document}